\documentclass[preprint, showpacs, amsmath, amssymb, prb, 12pt]{revtex4}
\usepackage{graphicx}
\usepackage{dcolumn}
\usepackage{bm}
\usepackage{color}
\begin{document}
\title{Strongly Correlated Two-Photon Transport in One-Dimensional Waveguide Coupled to A Two-Level System}
\author{Jung-Tsung Shen}
\author{Shanhui Fan}
\email{shanhui@stanford.edu}
\affiliation{Ginzton Laboratory, Stanford University, Stanford, CA
94305}
\date{\today}
\begin{abstract}
We show that two-photon transport is strongly correlated in one-dimensional waveguide coupled to a two-level system. The exact S-matrix is constructed using a generalized Bethe-Ansatz technique. We show that the scattering eigenstates of this system include a two-photon bound state that passes through the two-level system as a composite single particle. Also, the two-level system can induce effective attractive or repulsive interactions in space for photons.  This general procedure can be applied to the Anderson model as well.
\end{abstract}
 \maketitle

Creating strong photon-photon interaction at few-photon level is of great interest for quantum information sciences. In atomic gases, such interaction can be accomplished either with systems exhibiting electromagnetically induced transparency (EIT)~\cite{Harris:1998, Imamoglu:1997}, or by reaching the strong-coupling regime of a two-level atom in a high-Q cavity~\cite{Birnbaum:2005}. However, in an on-chip, solid-state environment, which is crucial for practical applications, there have been significant challenges in implementing these concepts. For example, it is difficult to create the long-lifetime dark state, which is required for EIT effects, in most practical solid-state environments~\cite{Turukhin:2001}. While the strong-coupling regime has been reached by placing a quantum dot in a high-Q photonic crystal microcavity~\cite{Reithmaier:2004,Yoshie:2004a}, doing so requires very accurate tuning of both the electronic and optical resonances to ensure simultaneous spectral and spatial overlaps~\cite{Badolato:2005}.

\begin{figure}[thb]
\scalebox{0.8}{\includegraphics[width=\columnwidth]{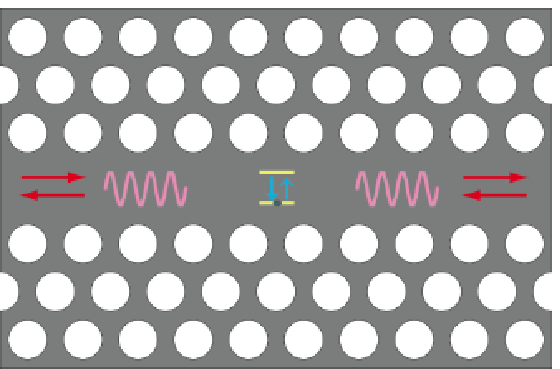}}
\caption{Schematics of the system. A two-level system is coupled to a one-dimensional continuum in which the photons, shown as wiggly waves, propagate in each direction.}\label{Fi:Geometry}
\end{figure}

In this Letter we propose and analyze in details an alternative scheme to create strong photon-photon interaction. Our approach exploits a unique one-dimensional feature for photon states in many nano-photonic structures. In a photonic crystal with a complete photonic band gap, for example, a line-defect waveguide forms a true one-dimensional continuum for photons, since there is no other states within the gap. Here we show that by coupling a two-level system to such continuum, strong photon-photon interactions can be created  (Fig.~\ref{Fi:Geometry}). (Below we refer to the two-level system as the ``atom''). In this system, the strong interaction arises from the fact that in a one-dimensional system, the re-emitted and scattered waves from the atom inevitably interfere with the incident waves. Moreover, since the atom, intuitively speaking, can at most absorb only one-photon at a time, the transport properties of multi-photon are strongly correlated.

Compared with previous solid-state approaches, our scheme does not require the presence of long-lifetime dark state. Neither does this scheme necessitate detailed spectral tuning or spatial control of the two-level system, since it operates in the weak-coupling regime, and thus the one-dimensional continuum can be broadband. Moreover, the Hamiltonian of the system actually describes an exact photonic analogue of the Kondo effect, which is important for processing electronic quantum bits~\cite{Zumbuhl:2004}. Our approach may therefore open a new avenue towards practical photon-based quantum information processing on-chip.

The system in Fig.~\ref{Fi:Geometry} is modeled by the Hamiltonian~\cite{Shen:2005,Shen:2005a}:
\begin{align}\label{E:hamiltonian}
H&=\int dx \left\{-i v_g c_{R}^{\dagger}(x)\frac{\partial}{\partial
x} c_{R}(x)
+ i v_g c_{L}^{\dagger}(x)\frac{\partial}{\partial x} c_{L}(x)\right. \notag\\
&\quad \left.+ \bar{V} \delta(x) \left(c^{\dagger}_{R}(x) \sigma_{-} +
c_{R}(x) \sigma_{+}
+ c^{\dagger}_{L}(x) \sigma_{-} + c_{L}(x) \sigma_{+}\right)\right\} \notag\\
&\quad +E_{e} a^{\dagger}_{e} a_{e}+ E_{g} a^{\dagger}_{g} a_{g}
\end{align}
where $v_g$ is the group velocity of the photons, and
$c_{R}^{\dagger}(x)$($c_{L}^{\dagger}(x)$) is a bosonic operator
creating a right-going(left-going) photon at $x$. $\bar{V}$ is the coupling constant,
$a^{\dagger}_{g}$($a^{\dagger}_{e}$) is the creation operator of the
ground (excited) state of the atom, $\sigma_{+}=a^{\dagger}_{e}
a_{g}$($\sigma_{-}=a^{\dagger}_{g} a_{e}$) is the atomic raising (lowering) ladder
operator satisfying $\sigma_{+}|n,-\rangle =|n,+\rangle$ and $\sigma_{+}|n,+\rangle =0$, where $|n, \pm\rangle$ describes the state of the system with $n$ photons and the atom in the excited ($+$) or ground ($-$) state. $E_{e}-E_{g}
(\equiv\Omega)$ is the transition energy. This Hamiltonian describes the situation where the propagating photons can run in both directions, and is referred to as ``two-mode'' model.

By employing the following transformation $c^{\dagger}_e (x)\equiv \frac{1}{\sqrt{2}}(c^{\dagger}_{R}(x)+ c^{\dagger}_{L}(-x))$, $c^{\dagger}_o (x)\equiv
\frac{1}{\sqrt{2}}(c^{\dagger}_{R}(x)- c^{\dagger}_{L}(-x))$, the original Hamiltonian is transformed into two decoupled ``one-mode''
Hamiltonians, \emph{i.e.,} $H=H_e + H_o$, where
\begin{align}
H_e &= \int dx (-i) v_g c_{e}^{\dagger}(x)\frac{\partial}{\partial x}
c_{e}(x) + \int dx V \delta(x)\left(c^{\dagger}_{e}(x)
\sigma_{-} + c_{e}(x) \sigma_{+}\right)+E_{e} a^{\dagger}_{e} a_{e}+
E_{g} a^{\dagger}_{g}
a_{g}\notag\\
H_o &= \int dx (-i) v_g c_{o}^{\dagger}(x)\frac{\partial}{\partial x}
c_{o}(x).
\end{align}$H_o$ is an interaction-free one-mode Hamiltonian, while
$H_e$ describes a non-trivial one-mode interacting model with coupling strength $V\equiv\sqrt{2} \bar{V}$. $H_e$ is identical in form to the s-d model~\cite{Anderson:1961,Wiegmann:1983}, which describes the S-wave scattering
of electrons off a magnetic impurity in three
dimensions. Here, however, instead of a fermionic operators describing electrons, we have bosonic operators describing photons.

The one-photon eigenstate for $H_e$ takes the form $|k\rangle \equiv \int dx [e^{i k x} \left(\theta(-x)+ t_k \theta(x)\right)c^{\dagger}(x) + e_k \sigma_{+}] |0, -\rangle$~\cite{Shen:2005,Shen:2005a}, where 
\begin{equation}
t_k \equiv\frac{k - \Omega - i \Gamma/2}{k -\Omega + i \Gamma/2}, \quad \Gamma\equiv V^2
\end{equation} is the transmission amplitude of magnitude 1, and $e_k = \frac{\sqrt{\Gamma}}{k-\Omega+i \Gamma/2}$ is the excitation amplitude. The single photon experiences resonance when its energy $k$ is close to the transition energy $\Omega$ of the atom. For notational simplicity, $v_g$ and $\hbar$ are set to 1, and the subscript ``$e$'' in $c^{\dagger}_e$ is dropped hereafter.

In this Letter we focus on the transport properties of the interacting Hamiltonian $H_e$ with two incident photons. For this Hamiltonian, as well as the Anderson model and the interacting resonance level model in condensed matter physics, attempts to diagonalize using the Bethe Ansatz have been published~\cite{Wiegmann:1983,Rupasov:1984a,Mehta:2006}. As we will emphasize below, however, a complete and correct description of the transport properties requires a careful re-examination of these solutions. In particular, the Bethe Ansatz solution constructed following the procedures in Ref.~[\onlinecite{Mehta:2006}] is in fact not complete for this purpose. Rather, to construct the scattering matrix, one needs \emph{one} additional two-photon bound state. These can all be derived by a systematic approach detailed below.

We first describe the general features of the scattering problem. Before and after the scattering, the photons are away from the atom, and thus the two-photon Hilbert spaces of the ``\emph{in}'' (before scattering) and ``\emph{out}'' (after scattering) states~\cite{Greiner:1996} are the same space of \emph{free} photons and consists of \emph{all} symmetric functions of the coordinates of the photons, $x_1$ and $x_2$. This Hilbert space is spanned by a complete basis $\{|S_{k,p}\rangle: k\leq p\}$ defined as
\begin{equation}
\langle x_1, x_2|S_{k, p}\rangle \equiv \frac{1}{2\pi}\frac{1}{\sqrt{2}}\left(e^{i k x_1} e^{i p x_2} +e^{i k x_2} e^{i p x_1}\right)= \frac{\sqrt{2}}{2\pi} e^{i E x_c}\cos\left(\Delta x\right), 
\end{equation}where $E=k+p$ is the total energy of the photon pair, $x_c \equiv 1/2(x_1 + x_2)$, $x\equiv x_1 - x_2$, and $\Delta \equiv (k - E/2) = 1/2 (k-p) \leq 0$.
Alternatively, the \emph{same} Hilbert space can instead be spanned by another basis $\{|A_{k,p}\rangle: k\leq p\}$ defined as
\begin{equation}
\langle x_1, x_2|A_{k,p}\rangle \equiv \frac{1}{2\pi}\frac{1}{\sqrt{2}}\,\mbox{sgn}(x)\left(e^{i k x_1} e^{i p x_2} -e^{i k x_2} e^{i p x_1}\right)
= \frac{\sqrt{2}i}{2\pi}\,\mbox{sgn}(x) \, e^{i E x_c}\sin\left(\Delta x\right)
\end{equation} where $\mbox{sgn}(x)\equiv \theta(x)-\theta(-x)$ is the sign function.  We emphasize that, while both $\{|S_{k,p}\rangle: k\leq p\}$ and $\{|A_{k,p}\rangle: k\leq p\}$ are complete~\cite{Schulz:1982}, arbitrary linear combination $\{ a_{k, p} |S_{k,p}\rangle + b_{k, p} |A_{k,p}\rangle: k\leq p\}$ may not be. 

The transport properties of two photons, in the presence of the atom, are described by the S-matrix ($\mathbf{S}$) that maps between the Hilbert space of the in and out states: $|\mbox{out}\rangle = \mathbf{S} |\mbox{in}\rangle$. The matrix element of the S-matrix, for example, $\langle S_{k, p}|\mathbf{S}|S_{k', p'}\rangle$ is the transition amplitude of the process~\cite{Greiner:1996}.

The S-matrix of the two-photon case, as will be derived below, can be diagonalized as 
\begin{equation}\label{E:SMatrix}
\mathbf{S}\equiv\sum_{k< p}  t_k t_p |W_{k,p}\rangle \langle W_{k, p}| + \sum_{E} t_E |B_E\rangle\langle B_E|, 
\end{equation} with
\begin{align}
|W_{k,p}\rangle &\equiv \frac{1}{\sqrt{(k-p)^2 +\Gamma^2}}\left[(k-p)|S_{k,p}\rangle + i \Gamma |A_{k,p}\rangle\right],\notag\\
\langle x_1, x_2 |B_E\rangle &\equiv \frac{\sqrt{\Gamma}}{\sqrt{4 \pi}} e^{i E x_c -\Gamma |x|/2}, \quad t_E \equiv \frac{E-2\Omega-2 i \Gamma}{E-2\Omega+ 2 i \Gamma}\label{E:Def}.
\end{align}

Now we prove Eqs.~(\ref{E:SMatrix})-(\ref{E:Def}) by first showing that $|W_{k,p}\rangle$ and $|B_{E}\rangle$ are eigenstates of the scattering matrix. A two-photon eigenstate for $H_e$ has the general form:
\begin{equation}
|\Phi\rangle \equiv\left(\int dx_1 dx_2 \, g(x_1, x_2) c^{\dagger}(x_1) c^{\dagger}(x_2) + \int dx \, e(x) c^{\dagger}(x) \sigma_{+}\right)|0, -\rangle,
\end{equation}
where $e(x)$ is the probability amplitude of the atom in the excited state.
Due to the boson statistics, the wavefunction satisfies $g(x_1, x_2) = +g(x_2, x_1)$. ($g(x_1, x_2)$ is continuous on the line $x_1 = x_2$ for bosons.)

From $H_e |\Phi\rangle = E |\Phi\rangle$, we obtain the equations of motion:
\begin{align}
\left(-i \frac{\partial}{\partial x_1}  -i \frac{\partial}{\partial x_2} - E\right) g(x_1, x_2) &+ \frac{V}{2} \left(e(x_1) \delta(x_2) + e(x_2) \delta(x_1)\right) =0, \notag\\
\left(-i \frac{\partial}{\partial x} - E + \Omega\right) e(x) &+ V \left(g(0, x) + g(x,0)\right) = 0,    
\end{align}
where $g(0, x) = g(x, 0) \equiv 1/2 \times(g(0^-, x) + g(0^+, x))$. The functions 
$g(x_1, x_2)$ and $e(x)$ are piecewise continuous. 


The interactions occur on the coordinate axes: $x_1=0$, and $x_2=0$. Applying the equations of motions gives the following boundary conditions on the boundary of quadrants II ($x_1< 0 < x_2$) and III ($x_1, x_2 < 0$):
\begin{align}\label{E:1}
-i \left(g(x_1, 0^+) - g(x_1, 0^-)\right)  &+\frac{V}{2} e(x_1)=0,\notag\\
\left(-i\frac{\partial}{\partial x_1}-(E-\Omega)\right) e(x_1) &+  V(g(x_1, 0^+) + g(x_1, 0^-))=0,
\end{align}and on the boundary of quadrants II ($x_1< 0 < x_2$) and I ($0 < x_1, x_2$):
\begin{align}\label{E:2}
-i \left(g(0^+, x_2) - g(0^-, x_2)\right)  &+\frac{V}{2} e(x_2)=0,\notag\\
\left(-i\frac{\partial}{\partial x_2}-(E-\Omega)\right) e(x_2) &+  V(g(0^+, x_2) + g(0^-, x_2))=0.
\end{align}These boundary conditions must be supplemented by a further condition\begin{equation}\label{E:3}
e(0^-) = e(0^+),
\end{equation}which ensures the self-consistency. 


By boson symmetry we only need to consider the half space $x_1 \leq x_2$. In this half space, suppose $g(x_1, x_2) =B_3 e^{i k x_1 + i p x_2} + A_3 e^{i p x_1 + i k x_2}$ for $x_1 < x_2 <0$, using Eqs.~(\ref{E:1}-\ref{E:3}),  we obtain $g(x_1, x_2) = t_k t_p (B_3 e^{i k x_1 + i p x_2} + A_3 e^{i p x_1 + i k x_2})$ for $0< x_1 <x_2$, provided $B_3/A_3 = (k - p - i \Gamma)/(k - p + i \Gamma)$ as required from the continuity condition of $e(x)$. Therefore, in the \emph{full} quadrant III, the in-state, $|W_{k,p}\rangle$ as defined by
\begin{align}
\langle x_1, x_2|W_{k,p}\rangle &=\left(A_3 e^{i k x_1 + i p x_2} + B_3 e^{i p x_1 + i k x_2}\right) \theta(x_1 - x_2)+ \left(B_3 e^{i k x_1 + i p x_2} + A_3 e^{i p x_1 + i k x_2}\right) \theta(x_2 - x_1)\notag\\
&\propto (k-p)\langle x_1, x_2 |S_{k, p}\rangle + i \Gamma \langle x_1, x_2|A_{k, p}\rangle,
\end{align}is an eigenstate of the S-matrix with eigenvalue $t_k t_p$. This construction and the form of the solution is in essence the Bethe Ansatz method~\cite{Wiegmann:1983, Mehta:2006}.  

The set $\{|W_{k,p}\rangle: k < p\}$ however does not form a complete set of basis of the free two-photon Hilbert space. Instead, there exists \emph{one} additional eigenstate of S-matrix, $|B_{E}\rangle$, defined by Eq.~(\ref{E:Def}). To see that $|B_{E}\rangle$ is an eigenstate of the S-matrix, suppose $g(x_1, x_2) = e^{i E x_c} e^{- \Gamma |x|/2}$ in quadrant III, again using Eqs.~(\ref{E:1}-\ref{E:3}), we obtain $g(x_1, x_2) = t_E e^{i E x_c} e^{- \Gamma |x|/2}$ in quadrant I.  Such bound state is important when calculating the ground-state energy in the Anderson model~\cite{Kawakami:1981}. We show here that it is also crucial to the scattering and transport properties. 

The set of eigenstates $\{|W_{k,p}, |B_{E}\rangle\}$ forms a complete and orthonormal basis that spans the free two-photon Hilbert space. The orthonormality check is straightforward: $\langle W_{k', p'}|W_{k,p}\rangle = \delta(k-k')\delta(p-p') = \delta(\Delta-\Delta')\delta(E-E')$, $\langle B_E|B_E\rangle = \delta(E-E')$, and $\langle W_{k, p}|B_{E}\rangle =0$. The completeness can be proven by checking that 
\begin{equation}
\mathbf{W}\equiv\sum_{k< p}  |W_{k, p}\rangle\langle W_{k, p}| + \sum_{E}|B_E\rangle\langle B_E|, 
\end{equation}
is indeed an identity operator.
This, together with the eigenvalues $t_k t_p$ and $t_E$, prove Eq.~(\ref{E:SMatrix}).


We note that the two-photon bound state described by $|B_{E}\rangle$, of which the spatial extent is $1/\Gamma$, behaves as an effective single composite particle with an energy $k+p$, and remains integral when passing through the atom. The two-level system therefore provides the capability of manipulating composite particles of photons~\cite{Jacobson:1995} without destroying them. This capability is important in quantum cryptography~\cite{Ekert:1991} and quantum lithography~\cite{Boto:2000}.

For an arbitrary in-state of $|\mbox{in}\rangle = |S_{k_1, p_1}\rangle$, the momenta distribution of the out-state $\langle S_{k_2, p_2} |\mbox{out}\rangle$ is 
\begin{equation}\label{E:SMatrixElement}
\langle S_{k_2, p_2}|\mathbf{S}|S_{k_1, p_1}\rangle =  t_{k_1}t_{p_1}\delta(\Delta_1 -\Delta_2)\delta(E_1 - E_2) + t_{k_1} t_{p_1}\delta(\Delta_1 +\Delta_2)\delta(E_1 - E_2)+ B\delta(E_1 -E_2)\end{equation}
where the first two terms 
are the direct and exchange terms of each individual incident momentum; the third term 
with
\begin{align}
B &=
\frac{16 i \Gamma^2}{\pi}\frac{E_1-2\Omega + i\Gamma}{\left[4\Delta_1^2 -(E_1 - 2\Omega + i\Gamma)^2\right] \left[4\Delta_2^2 -(E_1 - 2\Omega + i\Gamma)^2\right]}.
\end{align} represents the background fluorescence due to the scattering. 
When $\Delta_1 \neq \Delta_2$, $|B(E_1, \Delta_1, \Delta_2)|^2$ is the probability density for the outgoing photon pair in $(E_1, \Delta_2)$ state, when the incoming photon pair is in $(E_1, \Delta_1)$ state. 

The emergence of the background fluorescence is completely different from the well-known resonance fluorescence phenomenon where a strong laser beam is scattering off an ensemble of two-level systems~\cite{Scully:1997}. In the current two-photon case, the background fluorescence results from the fact that the momentum of each photon is not  conserved. Consequently the interactions with the two-level system redistribute the momenta of the photons over a continuous range, under the total energy and momentum conservation constraint. Furthermore, the locations of the poles in $B$, at 
$k_{1,2}=p_{1,2}=\Omega -i\Gamma/2$,
correspond approximately to either one of the photons having an energy at $\Omega$. Thus, the background fluorescence arises as one photon inelastically scatters off a composite transient object formed by the atom absorbing the other photon.

\begin{figure}[thb]
\scalebox{1}{\includegraphics[width=\columnwidth]{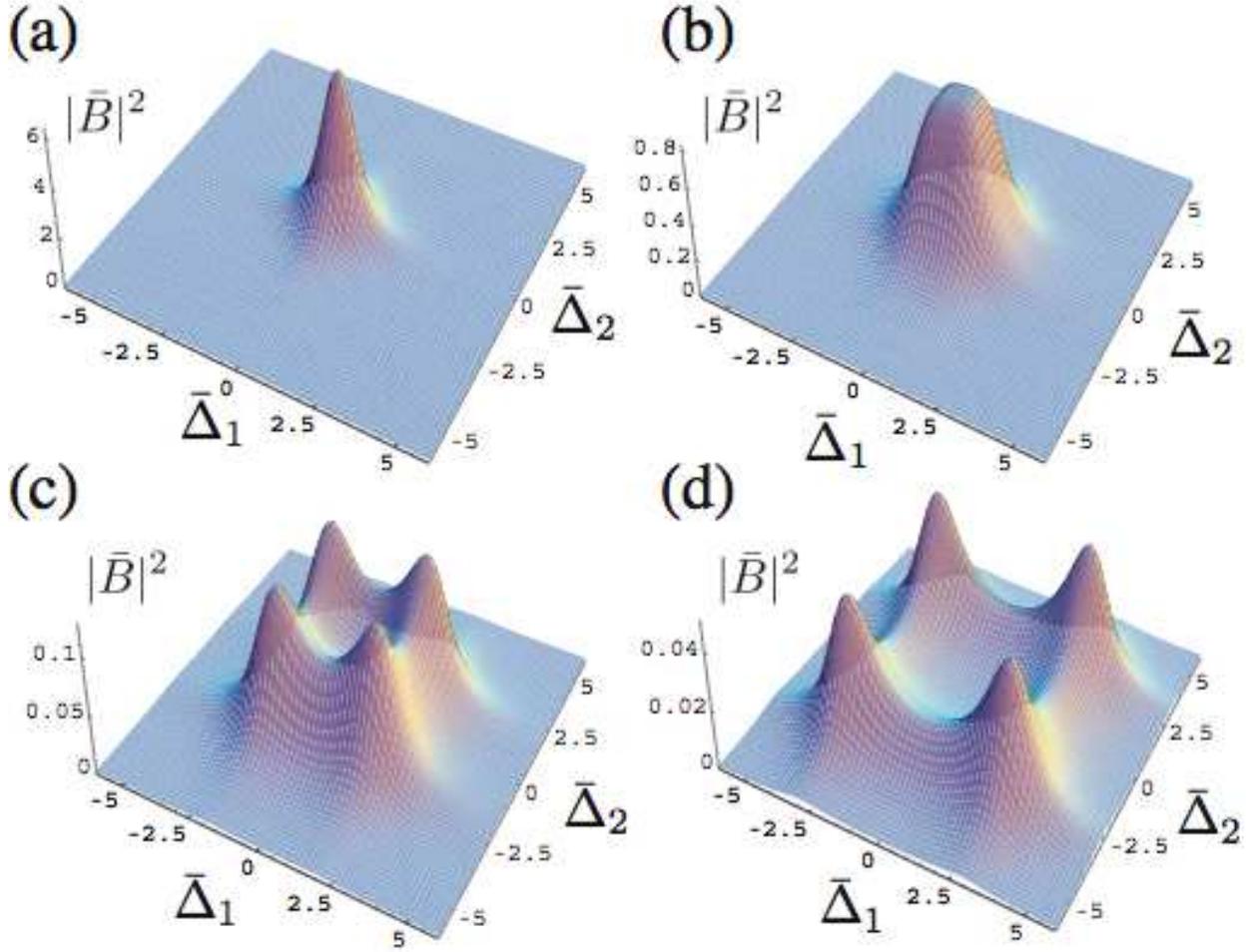}}
\caption{Background fluorescence as a function of $\bar{\Delta}_1$ and $\bar{\Delta}_2$ at various energy. (a) $\bar{E}=0$. (b) $\bar{E}=2$. (c) $\bar{E}=4$. (d) $\bar{E}=6$. $\bar{B}\equiv (\Gamma/2) B$, $\bar{E}\equiv (E-2\Omega)/(\Gamma/2)$, and $\bar{\Delta}\equiv\Delta/(\Gamma/2)$. For any given $E$, the in- and out-states can be completely specified by one quadrant in the $\Delta_1$-$\Delta_2$ plane.}\label{Fi:Background_3D}
\end{figure}

Fig.~\ref{Fi:Background_3D} plots normalized $|B(E, \Delta_1, \Delta_2)|^2$ for various photon pair energy $E$. $|B(E, \Delta_1, \Delta_2)|^2$ is an even function of $E-2\Omega$.
When $|E-2\Omega|\leq \Gamma$, there is a single peak centered at $\Delta_1=\Delta_2=0$. The height of the peak reaches maximum at $E=2\Omega$ (Fig.~\ref{Fi:Background_3D}(a)), and gradually decreases as $|E-2\Omega|$ increases. When $|E-2\Omega| = \Gamma$, the top of the peak becomes flat (Fig.~\ref{Fi:Background_3D}(b)). When  $|E-2\Omega| > \Gamma$, there are four peaks centered at $(\pm\sqrt{(E-2\Omega)^2-\Gamma^2}/2, \pm\sqrt{(E-2\Omega)^2-\Gamma^2}/2)$, respectively (Fig.~\ref{Fi:Background_3D}(c) and (d)). For any $E$ and $\Delta_1$, the locations of the peaks for $|B(E, \Delta_1, \Delta_2)|^2$ are independent of $\Delta_1$. In contrast, the $\delta$-functions in the S-matrix (Eq.~(\ref{E:SMatrixElement})) are located on the $\Delta_1=\Delta_2$ line.



\begin{figure}[thb]
\scalebox{1}{\includegraphics[width=\columnwidth]{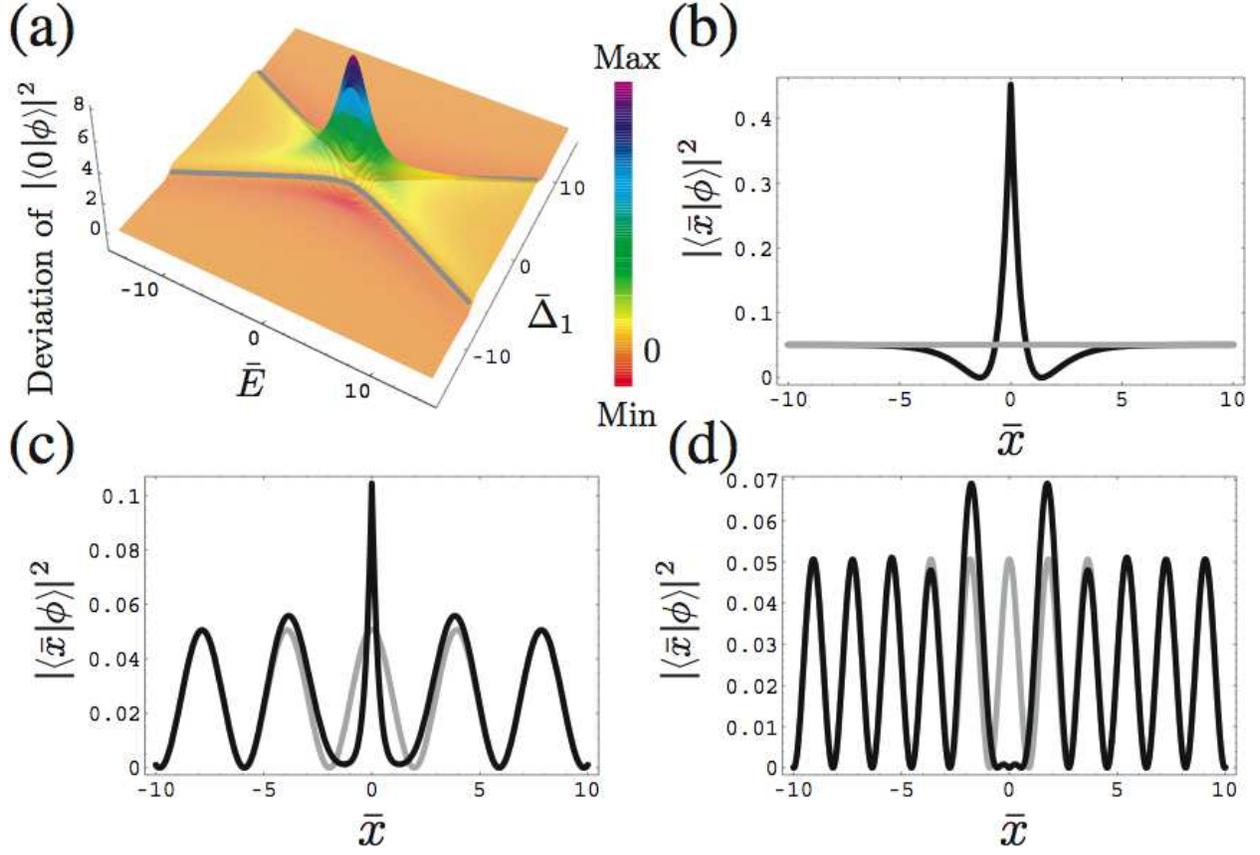}}
\caption{(a) Plot of the normalized deviation of the relative wavefunction: $|\langle \bar{x}=0|\phi\rangle|^2/(\sqrt{2}/2\pi)^2 - 1$. $\bar{x}\equiv (\Gamma/2) x$. The two gray lines indicates the zero value. (b)(c)(d): plots of $|\langle \bar{x}|\phi\rangle|^2$ for $\bar{E}_1=0$. (b) $\bar{\Delta}_1=0$. (c) $\bar{\Delta}_1=-0.8$. (d) $\bar{\Delta}_1=-\sqrt{3}$. Gray lines indicate the interaction-free case.}\label{Fi:Bunching}
\end{figure}

The emergence of the background fluorescence also manifests as an effective spatial interaction between the photons. For an in-state $|\mbox{in}\rangle = |S_{E_1, \Delta_1}\rangle$, the out-state is 
\begin{equation}
\langle x_c, x |\mbox{out}\rangle= 
e^{i E_1 x_c}\frac{\sqrt{2}}{2\pi}\left(t_{k_1} t_{p_1} \cos\left(\Delta_1 x\right)-\frac{4\Gamma^2}{4\Delta_1^2 -(E_1-2\Omega +  i\Gamma)^2} e^{i (E_1-2\Omega) |x|/2 -\Gamma |x|/2}\right)
\end{equation}
which takes the form $e^{i E_1 x_c} \langle x|\phi\rangle$, where $\langle x |\phi\rangle$ is the wavefunction in the relative coordinate $x$.  
The deviation of the out-state wavefunctions from that of interaction-free case is large when $\Delta_1 \simeq \pm (E_1/2 -\Omega)$, i.e., when at least one of the incident photons is close to resonance. 
Fig.~\ref{Fi:Bunching}(a) plots the normalized deviation of $|\langle x=0|\phi\rangle|^2$ from the interaction-free case as a function of $E_1$ and $\Delta_1$. A positive (negative) deviation implies that  the two photons bunch (anti-bunch) after scattering. The hyperbola $4\Delta_1^2 -(E_1-2\Omega)^2 = \Gamma^2$ indicate where the deviation is zero, thereby separate the bunching and anti-bunching regions. The deviation reaches maximum at $E_1 -2\Omega=\Delta_1=0$, when both incident photons are on resonance with the atom. The wavefunction for this case is shown in Fig.~\ref{Fi:Bunching}(b), which exhibits the exponentially decaying feature in $x$. The two photons form a bound state after scattering, with half-width in space about $1/\Gamma$. When $E_1 -2\Omega$ is kept at zero, the height of the peak at $x=0$ decreases with increasing $|\Delta_1|$ (Fig.~\ref{Fi:Bunching}(c)(d)). Fig.~\ref{Fi:Bunching}(d) shows the case for $\Delta_1=-\sqrt{3}\Gamma/2$, where the peak at $x=0$ is completely depleted. Both bunching and anti-bunching behavior occur at other non-resonant $E_1$ and $\Delta_1$ as well but is generally weaker. The resonance thus can induce either an effective repulsion or attraction between two photons. 
%



This work is supported by a Packard Fellowship.


\pagebreak
\newpage

\pagebreak
\newpage



\pagebreak
\newpage

\end{document}